\documentclass[a4paper,11pt]{article}

\usepackage{amssymb}
\usepackage{graphicx}

\newcommand{\be}{\begin{equation}}
\newcommand{\ee}{\end{equation}}
\newcommand{\bea}{\begin{eqnarray}}
\newcommand{\eea}{\end{eqnarray}}
\newcommand{\ie}{{\sl i.e.\/}}

\title{Accelerating lattice QCD simulations with 2 flavors of staggered fermions 
on multiple GPUs using OpenACC - a first attempt}

\author{Sourendu Gupta\footnote{email:sgupta@theory.tifr.res.in}
\\
{\em Department of Theoretical Physics, 
Tata Institute of Fundamental Research,} \\
{\em Homi Bhabha Road, Colaba, Mumbai 400005} \\\\
Pushan Majumdar\footnote{email:tppm@iacs.res.in} 
\\
{\em Department of Theoretical Physics, 
Indian Association for the Cultivation of Science,} \\
{\em 2A \& 2B Raja S.C. Mullick Road, Jadavpur, Kolkata 700032}}

\begin{document}

\maketitle

\begin{abstract}
We present the results of an effort to accelerate a Rational Hybrid
Monte Carlo (RHMC) program for lattice quantum chromodynamics (QCD)
simulation for 2 flavors of staggered fermions on multiple Kepler K20X
GPUs distributed on different nodes of a Cray XC30.  We do not use CUDA
but adopt a higher level directive based programming approach using the
OpenACC platform. The lattice QCD algorithm is known to be bandwidth
bound; our timing results illustrate this clearly, and we discuss how
this limits the parallelization gains.  We achieve more than a factor
three speed-up compared to the CPU only MPI program.
\end{abstract}


\section{Introduction}

At length scales of about a femtometer ($10^{-15}$m), \ie, at
sub-nuclear scales, the building blocks of nature are Fermions called
quarks interacting among themselves by exchanging Bosons called gluons.
This is the strong interaction, which is responsible for the stability
of the proton and thus of matter itself.

Quantum Chromodynamics (QCD), which is the theory of strong
interactions, is a quantum field theory with a SU(3) gauge symmetry
group. The gauge symmetry ensures that all color charges, whether quarks
or gluons, interact with the same strength and transform into each other
following the rules of the SU(3) group. The strength of the interaction
is characterized by a coupling which is a dimensionless number.

At high probe energies the coupling is small and analytic calculations
can be carried out in QCD using an expansion in the small coupling
which is called perturbation theory. At lower energies, which we
are interested in, this coupling becomes strong so that perturbation
theory cannot be used. The only known systematic way of carrying out
computations at this energy scale is by formulating the theory on a
space-time lattice \cite{lqcd}.

Lattice QCD computations, particularly implementations of the
Rational Hybrid Monte Carlo (RHMC) algorithm, have been large users
of high performance computing resources. In this paper we discuss
parallelization of the full RHMC QCD code on mixed CPU/GPU nodes using MPI
with OpenACC as a tool for GPU coding.  Both these aspects are new. We
outline the formulation of the computational problem in Sections 2 and
3. In Section 4 we give the parallelization scheme which we utilize. A
review of the state of the art in putting QCD on GPUs is given in
Section 5. In Section 6 we give details of the implementation using
OpenACC. Section 7 has results on the performance of the code. A short
summary can be found in Section 8.  In an appendix we give a Mathematica
code which generates a high precision rational approximation to the
fourth root of a matrix.

\section{Defining lattice QCD}

In an Euclidean quantum theory one needs to compute expectation values of
physically observable quantities by performing a functional integral over
the quark fields (denoted by $\psi$ and $\overline\psi$) and gauge fields
(denoted $U$). The relative weight of each configuration is proportional
to $\exp[-S(\bar\psi,\psi,U)]$ where $S$ is called the action functional,
and will be defined shortly. The expectation value of an observable
$O$ is defined by the functional integral
\be\label{lat}
   \langle O\rangle = \frac1Z\int_{\psi,\overline\psi,U}
       O(\psi,\overline\psi,U) {\rm ~e}^{-S(\psi,\overline\psi,U)}.
\ee
The normalization $Z$ is called the partition function and is
given by the requirement that the identity operator has expectation
value of unity. We will restrict ourselves to the case where the action
functional is positive, and the expectation values may be obtained by
a Monte Carlo method \cite{text}.

\begin{figure}
\centerline{\includegraphics[width=0.4\columnwidth]{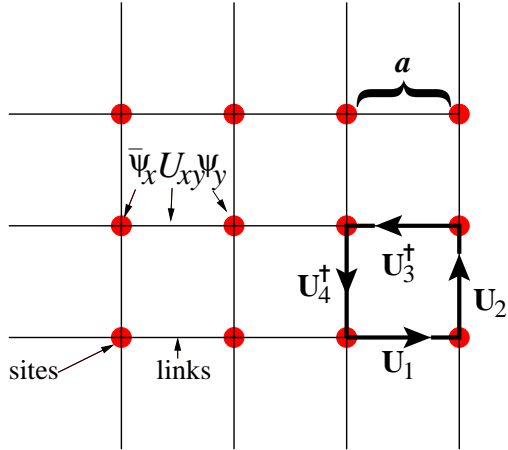}}
\caption{A 2-dimensional lattice. The matter fields $\bar\psi,\psi$ are
placed on sites while the gauge fields $U$ are placed on links. Here $a$
is the lattice spacing. As $a\rightarrow 0$ one recovers the continuum
theory. $\bar\psi_xU_{xy}\psi_y$ depicts the action of the Dirac
operator. The directed product $U_1U_2U_3^{\dagger}U_4^{\dagger}$ gives
us the $\Box$ in eq.\ (\ref{pfact}). }
\label{lattice}\end{figure}

Lattice QCD makes this possible by replacing the four-dimensional
space-time continuum by a hypercubic lattice of points, with lattice
spacing $a$ (see Fig. \ref{lattice}). Each point is now labeled
by four integers $(i_x,i_y,i_z,i_t)$. Each site may be given the address
$l=(((i_t-1)N_z+(i_z-1))N_y+(i_y-1))N_x+i_x$ where $N_x$ is the total
number of points in the $x$ direction, $N_y$ the total number of points in
the $y$ direction and so on. The volume of the lattice is $V=N_xN_yN_zN_t$.
Quark fields are placed on lattice sites and the gauge fields on the links
between the lattice sites. The gauge fields, $U$, are $3\times3$ (complex)
unitary matrices with determinant unity.  They require storage of $72V$
double precision words.  The quark fields are (anti-commuting) Grassman
variables. Since these have no efficient representation on computers,
they are integrated out. The part of the action which involves quark
fields is $S_f=\overline\psi M[U]\psi$, and $M[U]$ is a linear operator
involving the gauge fields $U$, called the Dirac operator. Integration
over the Dirac fields gives the weight
\be
   \int_{\bar\psi,\psi} {\rm e}^{-S_f} = \det M[U].
\ee
In the next section we will define the Dirac operator, and also explain
a trick for replacing the determinant by the positive definite quantity
$\det M^\dag[U]M[U]$. Although the Dirac operator is sparse, there is
no efficient algorithm for the computation of the determinant. The
best work-around turns out to be the introduction of a local Boson
(pseudo-Fermion) field $\phi$, which allows us to write
\be
  \det M^\dag[U]M[U] = \int_\phi {\rm e}^{-\phi^\dag(M^\dag M)^{-1}\phi}.
\ee
The pseudo-Fermion fields are three component complex fields and hence
require storage of $6V$ double precision words. The inversion
of the sparse positive definite operator $M^\dag M$ on the pseudo-fermion 
field $\phi$ is easily accomplished
by some version of a conjugate gradient algorithm.

The functional integral in eq.\ (\ref{lat}) can now be replaced by
\be\label{pflat}
   \langle O\rangle = \frac1Z\int_{\phi,U}O(U) \exp^{-S},
\ee
with the action
\be\label{pfact}
   S=\phi^\dag(M^\dag[U]M[U])^{-\alpha}\phi
      +\frac1{6g^2}\sum_{\mu,\nu}(3-{\rm tr\/}~\Box_{\mu\nu}),
\ee
and $\mu$ and $\nu$ are (non-parallel) directions on the lattice, and
$\Box$ is defined in Fig. \ref{lattice}.  The dimensionless parameter
$g$ is called the bare coupling of QCD.  The exact definition of $M[U]$
determines the value of $\alpha$ needed. The quantity 
$\alpha$ is related to the number of fermionic species under consideration;
our choice is given in Section 3.2.  Evaluation of the second term
in $S$ would take time linear in $V$. The matrix $M^\dag M$ has size of
order $V^2$, but is so sparse that it has order $V$ non-zero entries. As
a result, evaluating the first term would also take time of order $V$.
The time to obtain decorrelated measurements would take typically time
of order $V$, making the cost of such an algorithm of order $V^2$.

\section{Computing lattice QCD}

\subsection{The Algorithm}

To generate configurations of $U$ and $\phi$ distributed according to
eq.\ (\ref{pflat}), the algorithm of choice is known as Hybrid Monte Carlo
(HMC) \cite{hmc}, since this scales in CPU time as $V^{5/4}$.
HMC consists of three steps---
\begin{enumerate}
\item Introduce a random Gaussian noise $p$ (conjugate
to $U$) without affecting the correlation between
$\phi$ and $U$. Then $\exp[-S]$ in eq.\ (\ref{pflat}) goes over to
\be
   \exp[-H], \qquad H=\frac12p^2+S.
\ee
To create $\phi$ according to the correct distribution, generate another
Gaussian random vector $\xi$ and set 
\be
\phi=\left(M^\dag M\right)^{\alpha/2}\xi.
\ee
\item Using $H$ as a Hamiltonian, obtain the molecular dynamics
(MD) equations for $p$ and $U$.  The MD equation for U is given
by $\dot{U}=ipU$ while the one for p is given by $\dot H=0$
where the dot denotes derivative with respect to simulation-time.
Integrate the molecular dynamics equations obtained from the Hamiltonian
for a certain number of steps (simulation-time steps) to obtain a proposed
new configuration of $U(\tau)$ and $p(\tau)$.
\item Accept the proposed configuration $U(\tau),p(\tau)$ according to 
$$ P_{\rm acc}=\min\left\{ 1,\frac{\exp{[-H(U(\tau),p(\tau))]}}
{\exp{[-H(U(0),p(0))]}}
\right\}$$ where $(U(0),p(0))$ is the configuration in step 1.

\end{enumerate}
The steps $1-3$ constitute a so-called trajectory. The simulation consists of 
going over several thousands of trajectories and storing the configurations 
of $U$ generated. 

\begin{figure}
\centerline{\includegraphics[width=0.4\columnwidth]{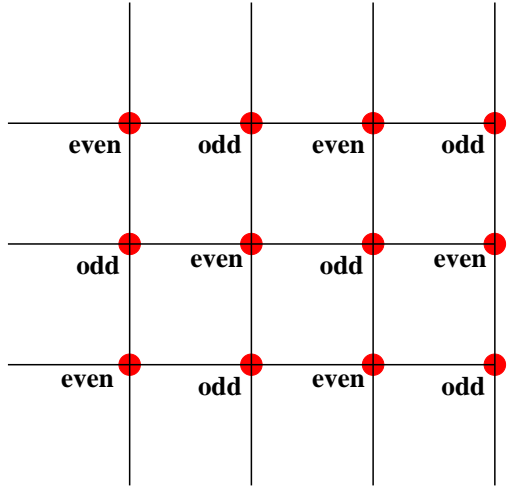}}
\caption{Even-odd decomposition of sites of a 2-dimensional lattice.}
\label{evenodd}\end{figure}

\subsection{Even-odd decomposition of the staggered fermion matrix}

Our primary aim is to do lattice QCD at finite temperature with two
light dynamical quarks and the formulation ideally suited for it is
called staggered Fermions with $\alpha=\frac{1}{2}$. There are some
special properties of this formulation which has a direct impact on the
computational problem.

The staggered Fermion matrix (where certain phases have been absorbed into
the $U$'s) is given by
\begin{eqnarray*}
   M_{ik}&=&2m\,\delta_{ik}+\sum_{\mu}(U_{i,\mu}\delta_{i,k-\mu}-
   U^{\dag}_{i-\mu,\mu}\delta_{i,k+\mu}) \\
   M^{\dag}_{ik}&=&2m\,\delta_{ik}-\sum_{\mu}(U_{i,\mu}\delta_{i,k-\mu}-
   U^{\dag}_{i-\mu,\mu}\delta_{i,k+\mu})
\end{eqnarray*}
If a lattice site has address $l$, we call it even or odd depending on
whether $l$ is even or odd (see Fig. \ref{evenodd}).  Reordering the
lattice so that the even sites and odd sites are grouped together,
we can rewrite $M$ and $M^{\dag}$ as

$$M=\left ( \begin{array}{cc} m_{oo} & D_{oe} \\ D_{eo} & m_{ee} \end{array} \right ) 
\qquad M^{\dag}=\left ( \begin{array}{cc} m_{oo} & D^{\dagger}_{oe} \\ 
D^{\dagger}_{eo} & m_{ee} \end{array} \right )$$

From the explicit expressions, we can see that $D^{\dagger}_{oe}=-D_{oe}$ and
$D^{\dagger}_{eo}=-D_{eo}$.
Therefore $\det (M^{\dagger}M)$ is given by
$$ \det \left (M^{\dagger}M \right ) = \det  \left [ \left (
\begin{array}{cc} m_{oo} & D^{\dagger}_{oe} \\ D^{\dagger}_{eo} & m_{ee} \end{array} \right )
\left ( \begin{array}{cc} m_{oo} & D_{oe} \\ D_{eo} & m_{ee} \end{array} \right )\right ]
$$ $$\qquad\qquad= \det \left (\begin{array}{cc} m^2_{oo}+D^{\dagger}_{oe}D_{eo} & 0 \\
0 & m^2_{ee}+D^{\dagger}_{eo}D_{oe} \end{array} \right )$$
From the block diagonal form, we get, $\det \left (M^{\dagger}M \right )=
\det \left (M^{\dagger}M \right )_{oo} 
\det \left (M^{\dagger}M \right )_{ee}$ and $\det \left (M^{\dagger}M \right )_{oo} = 
\det \left (M^{\dagger}M \right )_{ee}$.
Thus finally $\det (M) = \sqrt{\det \left (M^{\dagger}M \right )}= 
\det \left (M^{\dagger}M \right )_{ee}$
This is known as the half lattice formulation of the staggered quarks \cite{hmd}.
This halves the computational load of applying $M^\dag M$ on $\phi$.

\section{Slicing for parallelization}

From the previous Section, it can be seen that there are three main
types of operations which need to be carried out in the simulation.
\begin{enumerate}
\item Generation of large arrays of random numbers for $p$ and $\phi$. 
\item Computation of $\left(M^\dag M\right)^{-\frac12}\phi$. 
\item Carrying out a Metropolis accept/reject step at the end of each
trajectory.
\end{enumerate}
While steps 1 and 3 have to be carried out only once per trajectory, step
2 needs to be carried out at each integration step of the MD equations
and typically there are about 100 integration steps per trajectory. Our
main aim therefore will be to speed-up step 2.

$\left(M^\dag M\right)^{-\frac12}$ and $\left(M^\dag M\right)^{\frac14}$ 
can be expressed as rational approximations in the form 
\be\label{ratapprx}
\left (M^{\dagger}M \right )^{\frac{1}{b}} = c_0 + \sum_k \frac{c_k}{\left (M^{\dagger}M \right )
+ d_k}.
\ee
There are known algorithms to compute $c_k$ and $d_k$ and we use the
MiniMax approximation function of Mathematica to compute them for both
$b=4$ and $b=-2$ (see the appendix for the corresponding mathematica notebook). 
The number of coefficients required to reach a certain
degree of approximation depends upon the smallest eigenvalue of $M^\dag M$
(since the largest is fixed) and for our lightest masses we use about 30
coefficients to achieve an error below $10^{-11}$.  This approximation
has the advantage that $\left (M^{\dagger}M \right )^{-\frac{1}{2}}\phi$
can be computed using a shifted conjugate gradient \cite{mcg} using the same number
of shifts as we have coefficients \cite{KC}.

\begin{figure}
\centerline{\includegraphics[width=0.5\columnwidth]{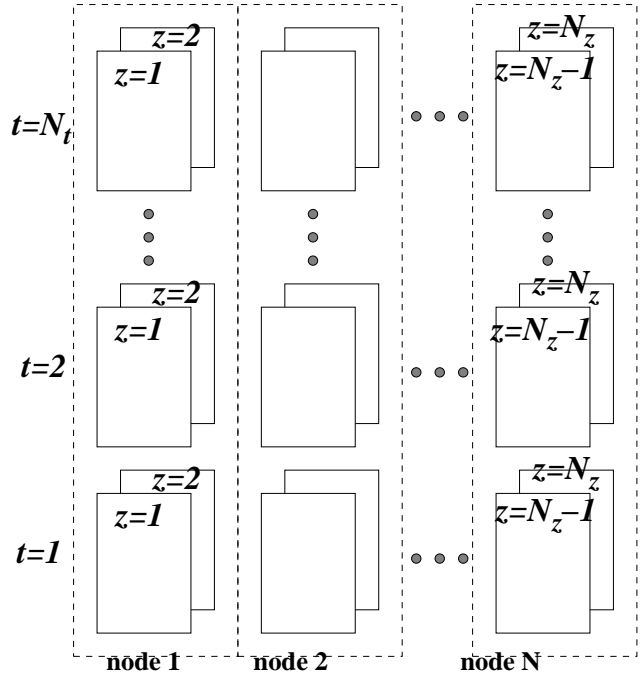}}
\caption{Maximal distribution of the lattice among different nodes. The
decomposition shown is along the $z$ direction, using only with two
z-slices on each node.}
\label{pardist}\end{figure}

The shifted conjugate gradient involves repeated application of $\left
(M^{\dagger}M \right )$ on vectors and this process is relatively
easily parallelizable. Currently the hardware which is ideally suited
for matrix-vector operations is the Graphics Processing Unit (GPU)
and this is the hardware of our choice for doing the simulations.

Unfortunately GPUs do not come with very large amounts of memory
(our hardware consists of Kepler GPUs with 6 GB of memory per card).
The lattice sizes we need to consider do not fit on a single GPU and we
are forced to split the lattice among multiple GPUs which are located on
different nodes. Also it is very inefficient to repeatedly copy data from
the CPU memory to GPU memory. We therefore have to divide the lattice
among different nodes in such a manner that the basic matrix-vector
operation can be done before one has to shift the data between the CPU
and GPU memories or between different nodes.

Finite temperature physics is simulated by shrinking one of the extents
of the lattice (typically $N_t$) depending on the temperature. Therefore
to maximize the number of nodes on which we can divide the lattice,
we choose to divide the lattice along the z direction. Our distribution
scheme is depicted in Fig. \ref{pardist}.

The application of $M^\dag M$ on a vector involves the application
of $D_{oe}$ and $D_{eo}$ in succession. In order to do this
efficiently we need to hold at least two z-slices on each node. At
the same time, we have to hold two more z-slices as boundaries in each
direction. Thus if we have $K$ z-slices for computation on each node
($K$ even and at least 2), we have to hold $K+4$ z-slices in
memory. Transferring the boundary data presents a significant bandwidth
bound for the code, as we show in Section 7.

\section{Lattice QCD on GPUs}

The first successful attempt to run lattice simulations on GPUs was
presented in \cite{video}. At that time programming GPUs required
lower level programming using OpenGL APIs. With the introduction of
CUDA \cite{cuda} (a C like high level language for programming GPUs) by
NVIDIA in 2007, GPUs started being adopted more and more for scientific
computing.  The initial results were quite encouraging and the cost
effectiveness of the GPUs was demonstrated \cite{lat09}. Shortly
afterwards a repository of CUDA codes for various types of lattice
QCD simulations (QUDA) was created and it is maintained and updated on
the QUDA website \cite{github}.  The performance of GPUs doubles and
quadruples if one uses single or half precision instead of the full
double precision. This led to exploration of mixed precision routines
in lattice QCD simulations \cite{lat10a}\cite{mixprec}.  Simulations on
multiple GPUs followed soon after. Initially it was multiple GPUs on a
single node \cite{lat10b} and then on GPUs on different nodes. It was
soon realized that putting only the conjugate gradient inverter on the
GPU was not enough and that one needed to put the whole trajectory on the
GPU \cite{lat10c} to minimize data transfer between the CPU and the GPU.

A slightly different direction was adopted in \cite{cusp} where the
authors used CUDA libraries to accelerate their lattice simulations.

CUDA was launched by NVIDIA and runs only on NVIDIA GPUs. A more general
programming language like OpenCL is needed for programming other GPUs
such as the Radeons manufactured by AMD. OpenCL packages for lattice
simulations were presented in \cite{qcdgpu} and \cite{cl2qcd}.

An object oriented approach to lattice QCD simulations on accelerators
has been under development for some time in Japan under the name Bridge++
\cite{bridge1}\cite{bridge2}\cite{bridge3}.

The performance of a full simulation code on GPUs using the Wilson
fermions with the clover improvement was reported in \cite{chroma}
using QUDA routines wherever possible.  Dirac operators with lattice
chiral symmetry were handled in \cite{overlap}

Pure Yang-Mills theory (i.e. without dynamical quark fields) simulation
codes have also been ported to GPUs \cite{bicudo}.

Lattice QCD simulation codes are bandwidth bound as one has to move
more data compared to the number of floating point operations on them
(i.e. FLOPS/Bytes $<$ 1). This ratio varies for different fermion
formulations \cite{pasi} and is the worst for the staggered fermion
formulation. On GPUs, the bandwidth available per compute core is much
lower than the CPU and this is the worst bottleneck to affect the
performance of the GPU program.  This problem is compounded for MPI
based multi-GPU codes as the GPU cannot launch MPI commands and the
data must be transferred back to the CPU for each MPI call.  We are
therefore addressing the most difficult parallelization problem and in
this article we examine carefully how the bandwidth limitation affects
our results and scaling.

Mixed precision routines are often used to boost performance of GPU
codes. Moreover since data movement is a bottleneck one often stores
and moves only two rows (or columns) of the ($3\times 3$) complex link
matrix and recovers the third by orthogonalization.  This is often known
as data compression.  It was observed that for light quark masses, mixed
precision routines often led to convergence problems for the multi-shift
conjugate gradient \cite{SD}. We therefore use fully double precision
routines. It was pointed out in \cite{pasi} that for single and half
precision data, compressed storage boosted performance significantly
while the effect was much less for double precision data. In our code
we have not yet used any data compression but hope to try it out soon.

\section{Implementation in OpenACC}

CUDA gives the programmer fine-grained control of the GPU, but it
also makes the code hardware specific and the programmer has to take
care of initializing the device and synchronizing it at the end of a
parallel region. Another option is to use a directive based programming
similar to OpenMP. OpenACC provides one such option. Our programs use
GPUs through OpenACC directives.  The recently introduced OpenMP 4.0
standard allows specification of accelerators and parts of the program
to be offloaded to the accelerators \cite{omp4} as part of the standard
OpenMP instruction set.

OpenACC is a programming standard for parallel computing on GPUs developed
by Cray, CAPS, NVIDIA and PGI.

The OpenACC Application Program Interface (API) lists a collection of
compiler directives to specify loops and regions of code to be offloaded
from a host CPU to an attached GPU, providing portability across operating
systems, host CPUs and GPUs. The directives allow programmers to create
a high-level CPU+GPU program without the need to explicitly initialize
the GPU or manage data and program transfers between the CPU and the
GPU \cite{openacc}.

OpenACC codes are similar to OpenMP codes in their directive structure. An
additional requirement is the creation of a data region which tells the
compiler when to move certain data items to the GPU or bring them back to
the CPU. The first attempts at using OpenACC for lattice QCD simulations
were reported in \cite{fermigpu} and \cite{bonati1}.  In this article
we concentrate mainly on multi-GPUs and we illustrate some (mainly
MPI) constructs below using snippets of our code in FORTRAN. For code
examples of a single node implementation see \cite{onenode}(FORTRAN)
and \cite{bonati2}(C).

A data region is created with the ACC data command and ended with a
ACC end data command as shown below.  The directive \verb|copy| means
that the variables should be copied into the GPU at the start of the
data region and copied out at its end. \verb|copyin| means it should be
copied into the GPU but need not be copied back to the CPU. Similar to
\verb|copyin|, there is also a \verb|copyout|. The directive \verb|create|
means the variable is defined on the GPU only. Only array variables need
to be declared in the ACC data declaration.\footnote{With the
advent of unified memory for GPUs, declaring separate data regions can
be avoided \cite{unimem}. Whether it will improve performance will need
to be tested.}

\begin{verbatim}
!$ACC data copy(nitcg,qg,ua)
!$ACC+ copyin(irsnd,ilsnd,ilrcv,irrcv,iblk)
!$ACC+ create(ap,rbndg,lbndg,qv,w1,fg)
!$ACC parallel loop present(irsnd,rbndg,ap)
         do i=1,ms*mt
           l=irsnd(i)
           rbndg(i,1)=ap(l,1)
           rbndg(i,2)=ap(l,2)
           rbndg(i,3)=ap(l,3)
         enddo
!$ACC update host(rbndg)
       call MPI_SENDRECV(rbndg,3*ms*mt,MPI_DOUBLE_COMPLEX,rneib,0,lbndg,
     1 3*ms*mt,MPI_DOUBLE_COMPLEX,lneib,0,MPI_COMM_WORLD, STATUS, ierr)
!$ACC update device(lbndg)
!$ACC parallel loop present(ilrcv,lbndg,ap) 
         do i=1,ms*mt
           l=ilrcv(i)
           ap(l,1)=lbndg(i,1)
           ap(l,2)=lbndg(i,2)
           ap(l,3)=lbndg(i,3)
         enddo
!ACC end data
\end{verbatim}
The blocking \verb| call MPI_SENDRECV| may be replaced by the equivalent 
non-blocking call \\ 
\verb|call MPI_Isend(rbnd,3*ms*mt,MPI_DOUBLE_COMPLEX,rneib,0,MPI_COMM_WORLD,req(1),ierr)|
\verb|call MPI_IRecv(lbnd,3*ms*mt,MPI_DOUBLE_COMPLEX,lneib,0,MPI_COMM_WORLD,req(2),ierr)|
\verb|call MPI_WAITALL(2,req,MPI_STATUSES_IGNORE,ierr)|\\
if necessary. ``\verb|req|" is an integer array of length 2.

Loops are parallelized by the directive parallel loop. If arrays are
referenced inside the loop, their status must be declared. They must
either be present on the GPU or be declared with \verb|present_or_copy|
attribute.  For multi-dimensional arrays, the order of the dimensions
are important. For example for the array ap, the ordering we used for GPU
operation was \verb|ap(V,3)| while for CPU operation we used the ordering 
\verb|ap(3,V)|
where \verb|V| is the lattice volume. There can be a substantial penalty in
performance if a non-optimal ordering is used. For a detailed discussion
see \cite{onenode}.  We also noticed that it was faster to use parallel
loops for vector operations rather than make BLAS calls. For reduction
operations too we found that parallel loops with the reduction variable
explicitly defined performed much better than the corresponding BLAS
function.  This holds for both Fermi and Kepler GPUs. For details see
\cite{onenode}.

MPI commands are launched from the CPU. So arrays used in the
communication must first be sent to the host using the \verb|update host|
directive. After the communication is over, the result must be sent to
the GPU using the \verb|update device| directive.  Due to these array
transfers between GPU and CPU, MPI calls from a GPU region of the code
are expensive and should be avoided as much as possible. The recently
introduced GPU-aware MPI and GPU Direct RDMA may cut down these overheads
but as our machine does not support them yet we were unable to test
their efficacy.

We need two sets of MPI calls repeatedly. Once when $D_{oe}$ and $D_{eo}$
are applied on the pseudofermion $\phi$ and once when the gauge fields
$U$ are updated in a MD integration step. The array involved in the
communication of the gauge fields is about 24 times the size of the array
required for communication of the boundary slices of the pseudofermion
and we found that each MPI call involving the boundary gauge fields
added about 1.5\% to the total execution time for our largest lattice.
The communication of gauge fields is needed right at the start of each
step of integration.  We found it was better to perform the communication
from the CPU and copy the entire data required for each single step
of the integration to the GPU. This was cheaper than the alternative,
which would have been to keep the whole data on the GPU for an entire
trajectory (100 integration steps) and copy the boundary data to the
CPU at each step of integration. An understanding of this
counter-intuitive observation could possibly be reached by profiling the
code. Unfortunately, we are unable to do this at the moment due to version
problems. We hope to come back to this issue in future. On the other
hand, for the single node program, where there is no MPI communication,
keeping the whole data on the GPU for the entire trajectory improves
performance significantly \cite{onenode}.

One more point worth mentioning is that in comparison to the older Fermi
GPUs which performed best with very light loops \cite{fermigpu}, on the
Kepler, loops with a heavier load seemed more beneficial. Our experience
was that on the Kepler GPU we gained by combining two light loops into
a slightly heavier single loop. This is probably because the computing
capacity of the individual Kepler cores is quite a bit higher than the
Fermi cores \cite{gpucomp}. We believe this trend has continued to the
Pascal and will continue further with the Volta GPU. It might therefore
be beneficial to revisit the load in individual loops as one updates
the GPU generation.

\section{Results}

In this Section we present the results of our test runs for different
lattice volumes and number of GPUs. All the tests were carried out
on a Cray XC30 which has a 2.8 Ghz 10 core (with 8 cores available
for computation) Intel Ivybridge processor and a K20X GPU per node.
The XC30 has three stages of interconnections. The nodes on a single
chassis have a backplane with a bandwidth of 14 Gbps. Different
chassis on a single cabinet are connected by copper cables again
with a bandwidth of 14 Gbps and finally the inter-cabinet connection
is through optical cables with a bandwidth of 12.5 Gbps \cite{xc30}.
This gives a nearest node point to point latency of $<1.4\mu s.$ We use
the Cray Fortran compiler for the Cray Programming Environment 2.5.5.
and link with the Cray scientific library for the CPU and the GPU
cray-libsci/16.07.1 and cray-libsci\_acc/16.03.1 respectively. The
MPI version we use is cray-mpich/7.4.1 and the CUDA toolkit version is
cudatoolkit/7.5.18-1.0502.10743.2.1.  Since lattice QCD programs are
bandwidth bound \cite{pasi}, it is useful to keep in mind that the CPU
memory bandwidth is about 32 GB/s \cite{xc30} while the GPU memory
bandwidth is about 180 GB/s \cite{gpuband}.

All our runs were carried out for the same input parameters viz. the quark
mass in lattice units was 0.002, the coupling $1/(6g^2)$ was taken to be
5.6. The number of conjugate gradient iterations varied between 5800 and
6600. These numbers are quite close to the actual runs that we do.  
We have discussed earlier, in Section 4, that when a lattice of size $N^4$
is split between $k$ nodes, so that each node works on $K=N/k$ slices,
then the number of lattice points on each node is actually $(K+4)N^3$.

\begin{table}
\caption{Comparing the performance of the CPU and GPU MPI program
with varying number of nodes ($k$) for a $32^4$ lattice.}
\begin{center}
\begin{tabular}{c|c|c|c}
\hline
$k$ & CPU time & GPU time & gain \\
\hline
2  & 32h45m58.86s & did not fit & $-$ \\
4  & 19h40m46.52s & 5h47m20.46s & 3.4 \\
8  & 13h14m13.25s & 4h4m53.60s & 3.24 \\
16 & 9h40m56.24s & 3h14m7.00s & 2.99 \\
\hline
\end{tabular}
\end{center}
\label{table:comp}
\end{table}

In Table \ref{table:comp} we compare the timings of identical runs on the
CPU and the GPU for a $32^4$ lattice on 2, 4, 8 and 16 nodes. We see that
the GPU runs are between 3 and 3.4 times faster than the CPU depending
on how big the lattice on a single node is. We obtain the performance
of our code in GigaFLOPS by instrumenting the CPU program with the
Cray performance analysis tools as reported in \cite{onenode}. The GPU
performance is obtained by scaling the CPU results by the timing ratio
between the CPU and GPU runs. This gives us about 30 GFLOPS/node for the
multi-shift conjugate gradient on a MPI program using 16 nodes. Single
node implementations in OpenACC reach about 40 GFLOPS for the multi-shift
conjugate gradient \cite{onenode} while single node CUDA implementations
can reach about 50 GFLOPS \cite{milc} on the K20X GPU.

Since bandwidth bounds lattice QCD program performance, it is of
interest to estimate the memory bandwidth obtained by individual OpenACC
kernels. This is difficult in a MPI program as data communication and
kernel execution can take place simultaneously. However, some idea
of the bandwidth usage can be obtained by examining the single node
performance for the program running a $16^4$ lattice using the Cray
performance analysis tools. We found that simple array operations like
multiplication of all elements by a scalar reached about 161 GB/s while inner 
product type kernels clocked 98 GB/s. Matrix-vector type operations like 
the Dirac operator $D_{oe}$ on a vector touched about 150 GB/s and the 
heaviest kernel which updates three large arrays and consumes about 60\% of 
the shifted conjugate gradient time returned a figure of about 90 GB/s.

\begin{figure*}
\centerline{\includegraphics[width=0.5\columnwidth,angle=-90]{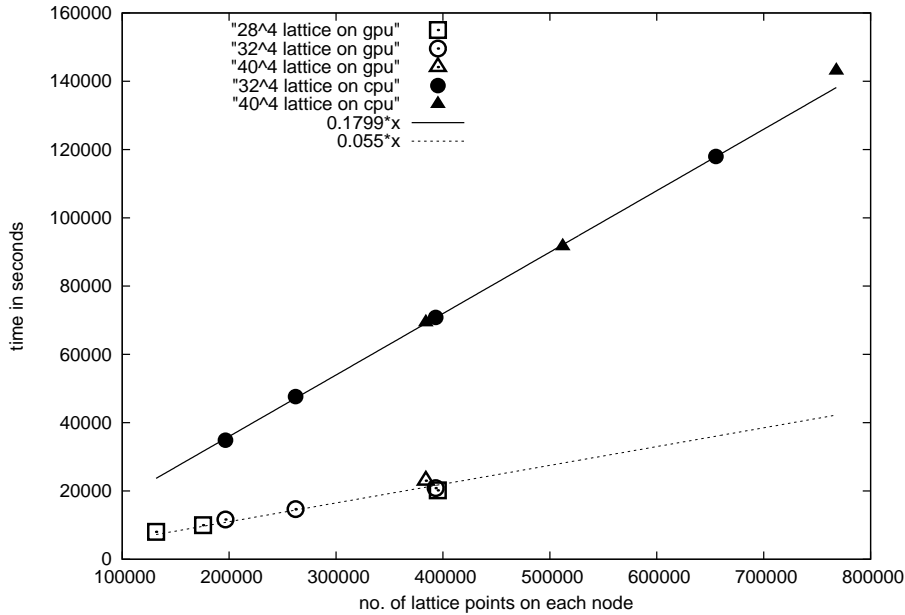}}
\caption{Volume scaling on various lattice sizes. The ratio of the slopes
of the two lines (which is about 3.27) gives the average speed-up on the GPU.}
\label{fig:vscale}
\end{figure*}

In Fig. \ref{fig:vscale} we plot the run timings of different lattice
volumes against the number of lattice points present on each node. The
timings for the different runs fall beautifully on two straight lines:
one for the CPU and one for the GPU. The ratio of the slopes of the two
lines (which is about 3.27) gives the average speed-up factor in going from
the CPU to the GPU. The straight lines are of the form $y=ax$ where $a$
is computed from one lattice volume (data point) for the CPU runs and one
lattice volume for the GPU runs. Except for the $40^4$ lattice on 5 CPU
nodes (with 768000 lattice points on each node), the rest of the points
just fall on the same straight line. The 5 node run for $40^4$ lattices
has a higher percentage of cache misses than the other smaller volumes.
We checked that the slope of the lines did not vary significantly if
any of the other lattice volumes were used to compute $a$.

\begin{table}
\caption{Slopes and intercepts for the straight lines describing the 
scaling of different lattice volumes on the CPU and the GPU as a function 
of $K$, the number of z-slices on each node.}
\begin{center}
\begin{tabular}{c|c|c}
\hline
description & slope ($a$) & intercept ($b$) \\
\hline
$32^4$ lattice on CPU & 5909  $\pm$ 46  & 23508 $\pm$ 427 \\
$40^4$ lattice on CPU & 12370 $\pm$ 425 & 43615 $\pm$ 2250 \\
$28^4$ lattice on GPU & 1015  $\pm$ 5   & 5974  $\pm$ 42 \\
$32^4$ lattice on GPU & 1533  $\pm$ 3   & 8574  $\pm$ 17 \\
\hline
\end{tabular}
\end{center}
\label{table:fits}
\end{table}

\begin{figure}
\centerline{\includegraphics[width=0.35\columnwidth,angle=-90]{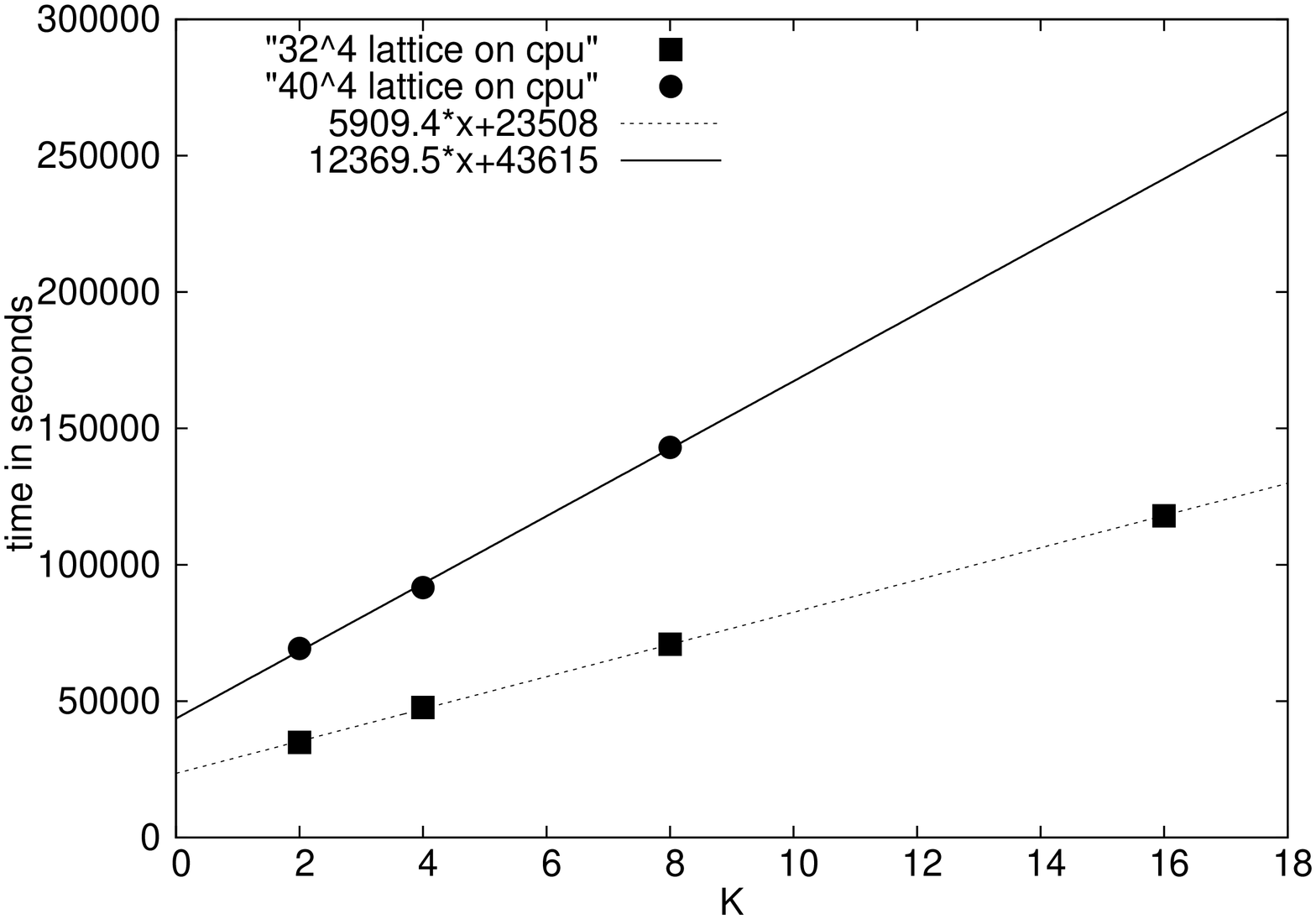}\hspace{5mm}
\includegraphics[width=0.35\columnwidth,angle=-90]{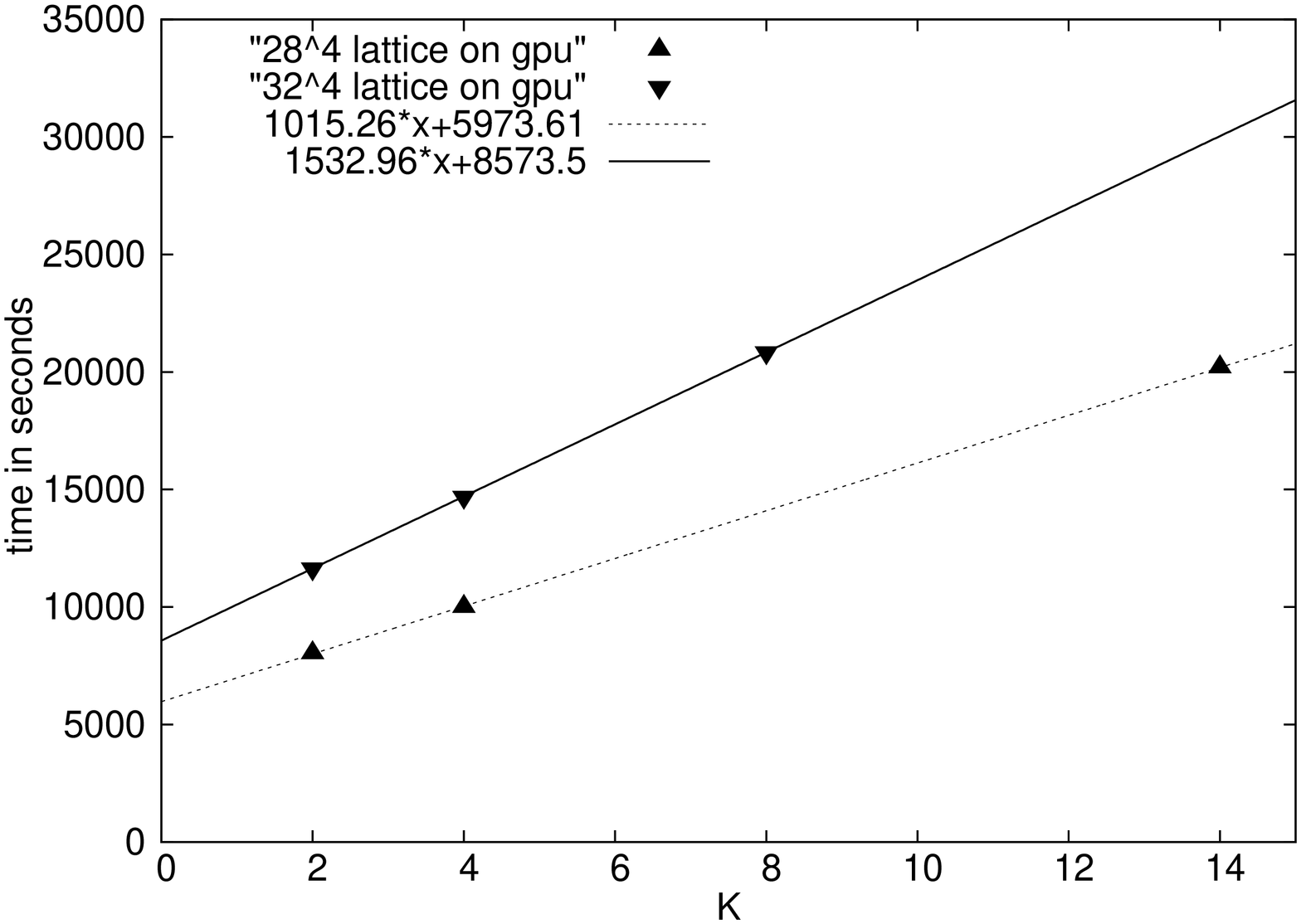}}
\caption{Wall-clock time against number of z-slices used in computation,
for two different lattice sizes. The intercept of the lines on the
$y$-axis give an estimate of the time required to move the boundary
slices from the memory to the CPU (left) or the GPU (right).}
\label{fig:bbound}
\end{figure}

We explore this feature in a bit more detail in Fig. \ref{fig:bbound}.
In this figure we plot the run times against $K$, the number of z-slices
on each node. The points are well described by a straight line of the
form $y=ax+b$. A linear fit to each of the lattice volumes on the CPU
and the GPU gives the slope $a$ and the intercept $b$ for the four lines
(see Table \ref{table:fits}). It seems odd that the time required
is finite even when $K=0$, \ie, there is no computation. However, from
our earlier discussion, it is clear that this time is needed to move the
boundary data between the memory and the CPU (left) or the GPU (right).

The ratios of the intercepts, $b$, must be the ratio of the number
of words transferred. On the CPU this ratio should be $(32/40)^3 =
0.512$. This is in agreement with the observed ratio of the intercepts,
which is $0.54\pm0.03$. On the GPU the ratio should be $(28/32)^3=0.67$.
The observed ratio of intercepts is $0.697\pm0.005$, which is close to,
but not in full agreement with the expected value. The little discrepancy
may be due to various reasons.  In the GPU computation there are transfers
between CPUs as well as between CPU and GPU, and the bandwidths for
these are different.  In addition, the asymmetry in data access time
for a core can be much higher in the GPU than in the CPU. Understanding
the slight discrepancy between measurement and expectation may need more
instrumentation of the code than was available to us.

The ratios of the slopes should be understood in terms of the amount
of computation performed. Numerically these should be the same ratios
as for the intercepts. On the CPU the ratio of slopes is $0.48\pm0.02$,
which is consistent with $0.512$ at the 95\% confidence level. This is
also the case on the GPUs, where the observed ratio is $0.662\pm0.004$
and the expected one is $0.67$.

\begin{figure}
\centerline{\includegraphics[width=0.4\columnwidth,angle=-90]{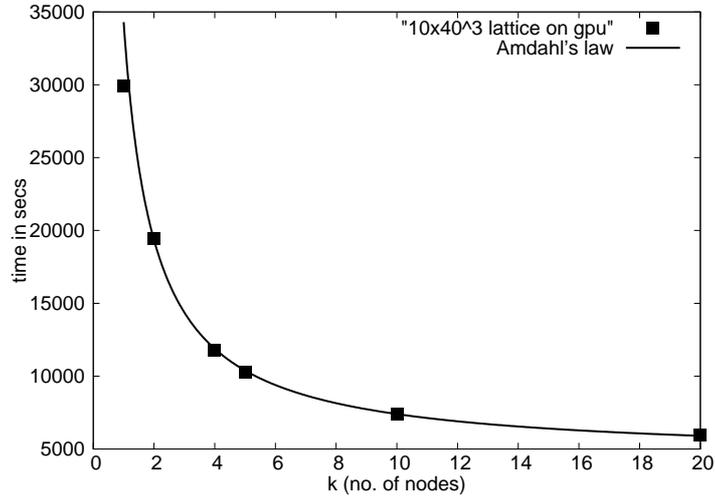}}
\caption{Strong scaling plot showing wall-clock time vs nodes for a
$10 \times 40^3$ lattice on the GPU. The points are well described by a
curve of the form $y=a/k+b$ with $a=29888\pm272$ and $b=4410\pm73$. The
one node result is from a non-MPI program.}
\label{fig:strongscaling}
\end{figure}

Finally in Fig. \ref{fig:strongscaling}, we show a strong scaling
plot for the lattice size $10\times 40^3$. The points are fitted well
by a curve of the form $a/k+b$ where $k$ stands for the number of GPUs
used. A non-zero value of $b$ is due to the non-parallelized part of
the code. From Amdahl's law we obtain the parallel speed up of any program
to be $s=1/[(1-P)+P/k]$ where $P$ is the fraction of parallel code in the
program and $k$ is the number of parallel threads. The fit shown in
Fig. \ref{fig:strongscaling} implies that $P$ is approximately 87\%.
We plan to examine the remaining 13\% in future to see whether some
fraction of this can also be parallelized. The single node point is a
non-MPI program. It is a coincidence that the single node run timing
lies practically on the MPI curve.  It was not used for obtaining $a$
and $b$.  Actually the single node program was slowed down slightly to
reduce its memory footprint as otherwise a $10\times 40^3$ lattice does
not fit on a single K20X card.

\section{Summary}

In this article we have described our implementation of a lattice QCD
simulation program using an RHMC algorithm on the Ivybridge-K20X (CPU-GPU)
heterogeneous architecture using the OpenACC programming paradigm. We
find the OpenACC platform easy to use and it offers the possibility of
a unified code that can run on both the CPU and the GPU once OpenACC
is included in the OpenMP standard. Of course, to gain efficiency, some
amount of data structure reorganization will be necessary as one migrates
from one hardware to another, but the programming barrier will
be much reduced.

Lattice QCD simulation programs are bandwidth bound. This is very
clearly seen in our scaling plots. One important aspect of the bandwidth
bound is that when slicing up the lattice for parallel handling a deep
boundary has to be kept. We used naive staggered quarks where the boundary
is two slices in each direction. ``Improved'' quarks often have deeper
boundaries, leading to further memory transfer bottlenecks.
It is possible that putting the boundary data in a separate 
array may alleviate such bottlenecks but that is a study for the future.

Although the theoretical maximum gain of a GPU accelerator is a
factor of about 5.6 over the CPU, we achieve a factor of $3.4$ in our runs. We
obtain about 30 GFLOPS/node for the multi-shift conjugate gradient on a
MPI program on 16 nodes.  This parallel performance is to be compared
to the nearly 40 GFLOPS which has been obtained using OpenACC
\cite{onenode} or 50 GFLOPS using CUDA \cite{milc} for similar multi-shift conjugate
gradients on single K20X GPUs.

\section*{Acknowledgments}
The authors would like to acknowledge useful discussions with Patricia Balle and 
Stephen Behling of Cray on measuring the memory bandwidth of individual OpenACC 
kernels.
 
\section*{Appendix : Mathematica code for the rational approximation}

Below we present a Mathematica code to compute the rational
approximation to the function $x^{\frac14}$ in the interval \verb|xmin|
and \verb|xmax|. \verb|xmin| and \verb|xmax| are given by bounds on the
smallest and largest eigenvalues of $M^{\dagger}M$ where $M$ is the
massive Dirac operator. For staggered fermions, in the convention we
use, \verb|xmin| has to be $<(2ma)^2$ and \verb|xmax| has to be $>32$.
Since the error of the approximation increases towards the lower or
upper ends, we choose the range of approximation to be slightly larger
than the spectral bound of the operator.

For the code snippet below $ma=0.005$ and so \verb|xmin| was chosen as
$5\times 10^{-5}$.  \verb|xmax| was always kept at 34. The degree of the
polynomials in the numerator and the denominator was chosen to be 24 each
so that the approximation is of the form given in eq.(\ref{ratapprx}).

\begin{verbatim}
<< FunctionApproximations`
xmin = 5*10^(-5) ;
xmax = 34 ;
degnum = 24 ;
degden = 24 ;
mmlist = MiniMaxApproximation[x^(1/4),{x,{xmin,xmax},degnum,degden},
   Bias -> -0.4, Brake -> {85,85}, MaxIterations -> 1000, WorkingPrecision -> 1000];
mmfunc = mmlist[[2,1]];
mf1 = Apart[mmfunc];
coeff0 = mf1[[1]];
mf2 = (mf1 - coeff0)[[2]];
rn = SetPrecision[Table[RandomReal[{1^(-2),34},24]], 80];
rhs = SetPrecision[Table[mf2/.x -> rn[[i]], {i,1,24}], 80];
rrts = SetPrecision[Roots[mf2[[2]]^(-1) == 0, x],80];
lhs = SetPrecision[Table[((x-rrts[[i,2]])/.x -> rn[[j]])^(-1),{j,24},{i,24}], 80];
coeff = Table[SetPrecision[LinearSolve[lhs,rhs], 80]];
mf = SetPrecision[coeff0 + Sum[coeff[[i]]/(x-rrts[[i,2]]),{i,1,24}], 16]
\end{verbatim}

This coefficients are in \verb|coeff0| and the table \verb|coeff| and
the shifts are in the table \verb|rrts|. This approximation has an error
of $< 10^{-12}$ throughout the approximation range. A higher degree of
approximation can be achieved by increasing the degree of the polynomials
in the numerator and the denominator.

An approximation may also be obtained by choosing the degree of the
numerator to be $n-1$ and the degree of the denominator as $n$. That
would set \verb|coeff0|=0. Then \verb|mf2=mmfunc|.  However we found
that keeping \verb|coeff0| halved the error of the approximation without
increasing the number of shifts required in the conjugate gradient.


\begin{thebibliography}{99}
\bibitem{lqcd}
  P. Weisz, P. Majumdar, Lattice gauge theories, Scholarpedia 7(4) (2012) 8615 
\verb|http://www.scholarpedia.org/article/Lattice_gauge_theories|
\bibitem{text}
  I.~Montvay, G.~M\"unster, Quantum Fields on a Lattice,
  Cambridge Monographs on Mathematical Physics, Cambridge University Press,
  Cambridge UK, (1994).
\bibitem{hmc}
  S.~Duane, A.~D.~Kennedy, B.~J.~Pendleton, D.~Roweth, Hybrid Monte Carlo, 
  Phys.\ Lett.\ B 195 (1987)  216.
\bibitem{hmd} 
  S. Gottlieb, W. Liu, D. Toussaint, R. L. Renken, R. L. Sugar, 
  Hybrid-molecular-dynamics algorithms for the numerical simulation of quantum chromodynamics,
  Phys.\ Rev.\ D 35 (1987) 2531.

\bibitem{mcg}
  B. Jegerlehner,
  Krylov space solvers for shifted linear systems,
  arXiv:hep-lat/9612014, 1996

\bibitem{KC}
  M.A. Clark, A.D. Kennedy, The RHMC algorithm for 2 flavors of dynamical staggered fermions, 
  Nucl. Phys. Proc. Suppl. 129 (2004) 850.
\bibitem{video} G.I. Egri, Z. Fodor, C. Hoelbling, S.D. Katz, D. Nogradi, K.K. Szabo,
 Lattice QCD as a video game, 
Comput. Phys. Commun. 177 (2007) 631.
\bibitem{cuda} \verb|https://developer.nvidia.com/cuda-zone|

\bibitem{lat09} M.A. Clark, 
QCD on GPUs: cost effective supercomputing,
PoS (LAT2009) (2009) 003. arXiv:0912.2268

\bibitem{github} \verb|https://github.com/lattice/quda|

\bibitem{lat10a} S. Gottlieb, A. Torok, V. Kindratenko, G. Shi, 
QUDA programming for staggered quarks, 
PoS (Lattice2010) (2010) 026.

\bibitem{mixprec} M.A. Clark, R. Babich, K. Barros, R.C. Brower, C. Rebbi,
Solving Lattice QCD systems of equations using mixed precision solvers on GPUs,
Comput. Phys. Commun. 181 (2010) 1517
  
\bibitem{lat10b} M. Hayakawa, K.-I. Ishikawa, Y. Osaki, S. Takeda, S. Uno, N. Yamada, 
Improving many flavor QCD simulations using multiple GPUs, 
PoS (Lattice2010) (2010) 325. arXiv:1009.5169
  
\bibitem{lat10c} C. Bonati, G. Cossu, M. D'Elia, A.D. Giacomo, 
Staggered fermions simulations on GPUs, 
PoS (Lattice2010) (2010) 324. arXiv:1010.5433
  
\bibitem{cusp} R. Galvez, G. van Anders, 
Accelerating the solution of families of shifted linear systems with CUDA, 
arXiv:1102.2143, (2011)

\bibitem{qcdgpu} V. Demchik, N. Kolomoyets,
QCDGPU: open-source package for Monte Carlo lattice simulations on OpenCL-compatible multi-GPU systems,
Presented at the Third International Conference ``High Performance Computing" (HPC-UA 2013), Kyiv,
Ukraine. arXiv:1310.7087

\bibitem{cl2qcd} O. Philipsen, C. Pinke, A. Sciarra, M. Bach,
CL2QCD-Lattice QCD based on OpenCL,
PoS (LATTICE2014) (2014) 038. arXiv:1411.5219

\bibitem{bridge1} S.Motoki, S.Aoki, T.Aoyama, K.Kanaya, H.Matsufuru, Y.Namekawa, H.Nemura, Y.Taniguchi, S.Ueda, N.Ukita,
Development of Lattice QCD Simulation Code Set ``Bridge++” on Accelerators,
Procedia Computer Science Volume 29 (2014) 1701.

\bibitem{bridge2} H.Matsufuru, S.Aoki, T.Aoyama, K.Kanaya, S.Motoki, Y.Namekawa, H.Nemura, Y.Taniguchi, S.Ueda, N.Ukita,
OpenCL vs OpenACC: Lessons from Development of Lattice QCD Simulation Code,
Procedia Computer Science Volume 51 (2015) 1313.

\bibitem{bridge3} S. Ueda, S. Aoki, T. Aoyama, K. Kanaya, H. Matsufuru, S. Motoki, Y. Namekawa, H. Nemura, Y. Taniguchi, N. Ukita,
Lattice QCD code ``Bridge++" on multi-thread and many core accelerators,
PoS (LATTICE2014) (2015) 036.

\bibitem{chroma} F. T. Winter, M. A. Clark, R. G. Edwards, B. Joo,
A framework for lattice QCD calculations on GPUs,
IEEE 28th International Parallel and Distributed Processing Symposium 2014.

\bibitem{overlap} A. Alexandru ,
Lattice Quantum Chromodynamics with Overlap Fermions on GPUs,
Computing in Science \& Engineering Volume: 17 Issue: 2 (2015) 14. 

\bibitem{bicudo} N. Cardoso, P. Bicudo, SU (2) lattice gauge theory simulations on Fermi GPUs, Comput. Phys. Commun. 184 (2013) 509.

\bibitem{pasi} R. Brower, Talk at PASI (Valpariso, Chile) Jan 2011, 
\verb|https://www.bu.edu/pasi/files/2011/05/Chile_browerPartIII.pdf|

\bibitem{SD} Private communication from Saumen Datta. 

\bibitem{omp4} See the construct target at \\ \verb|http://www.openmp.org/wp-content/uploads/OpenMP-4.5-1115-F-web.pdf|

\bibitem{openacc}  http://www.openacc-standard.org/

\bibitem{fermigpu}  P. Majumdar, Lattice Simulations using OpenACC compilers, PoS (LATTICE2013) (2014) 031.

\bibitem{bonati1} C. Bonati, E. Calore, S. Coscetti, M. D'elia,
Development of scientific software for hpc architectures using open acc: The case of lqcd, 
Software Engineering for High Performance Computing in Science (SE4HPCS), 2015. 

\bibitem{onenode} P. Majumdar, Lattice QCD simulations using the OpenACC platform, J.\ Phys.\ Conf.\ Ser. 759 no.1 (2016) 012070.

\bibitem{bonati2} C. Bonati, E. Calore, S. Coscetti, M. D'Elia, M. Mesiti,
F. Negro, S.F. Schifano, G. Silvi, R. Tripiccione,
Design and optimization of a portable LQCD Monte Carlo code using OpenACC
Int. J. Mod. Phys. C 28 (2017) 1750063. arXiv:1701.00426 

\bibitem{unimem} N. Sakharnykh,
Combine OpenACC and Unified Memory for Productivity and Performance, (2015)
\verb|https://devblogs.nvidia.com/parallelforall/|
\verb|combine-openacc-unified-memory-productivity-performance|

\bibitem{gpucomp} X. Mei, X. Chu, 
Dissecting GPU Memory Hierarchy through Microbenchmarking, (2015)
\verb|https://arxiv.org/pdf/1509.02308.pdf|

\bibitem{xc30} \verb|https://www.nersc.gov/assets/Uploads/NERSC.XC30.overview.pdf|

\bibitem{gpuband} E. Phillips, M. Fatica,
Optimizing the High Performance Conjugate Gradient Benchmark on GPUs, (2014)
\verb|https://devblogs.nvidia.com/parallelforall/|
\verb|optimizing-high-performance-conjugate-gradient-benchmark-gpus/|

\bibitem{milc} R. Li, C. DeTar, S. Gottlieb, D. Toussaint, 
MILC Code Performance on High End CPU and GPU Supercomputer Clusters, 
  Presentation in Lattice 2017 \verb|https://makondo.ugr.es/event/0/session/97/contribution/385|

\end{thebibliography}
\end{document}